\documentclass[aps,prl,twocolumn,groupedaddress]{revtex4}
\usepackage{graphicx}
\usepackage{amsmath}
\usepackage{amssymb}
\usepackage{bm}
\usepackage{times}

\newcommand{\ket}[1]{\ensuremath{\left|{#1}\right\rangle}}

\begin{document}

\title{Experimental Observation of Environment-induced Sudden Death of Entanglement}
\author{M. P. Almeida}
\affiliation{Instituto de F\'{\i}sica, Universidade Federal do Rio
de Janeiro, Caixa Postal 68528, Rio de Janeiro, RJ 21941-972,
Brazil}
\author{F. de Melo}
\affiliation{Instituto de F\'{\i}sica, Universidade Federal do Rio de
Janeiro, Caixa Postal 68528, Rio de Janeiro, RJ 21941-972, Brazil}
\author{M. Hor-Meyll}
\affiliation{Instituto de F\'{\i}sica, Universidade Federal do Rio
de Janeiro, Caixa Postal 68528, Rio de Janeiro, RJ 21941-972,
Brazil}
\author{A. Salles}
\affiliation{Instituto de F\'{\i}sica, Universidade Federal do Rio de
Janeiro, Caixa Postal 68528, Rio de Janeiro, RJ 21941-972, Brazil}
\author{S. P. Walborn}
\email[]{swalborn@if.ufrj.br}
\affiliation{Instituto de F\'{\i}sica, Universidade Federal do Rio
de Janeiro, Caixa Postal 68528, Rio de Janeiro, RJ 21941-972,
Brazil}
\author{P. H. Souto Ribeiro}
\affiliation{Instituto de F\'{\i}sica, Universidade Federal do Rio
de Janeiro, Caixa Postal 68528, Rio de Janeiro, RJ 21941-972,
Brazil}
\author{L. Davidovich}
\affiliation{Instituto de F\'{\i}sica, Universidade Federal do Rio de
Janeiro, Caixa Postal 68528, Rio de Janeiro, RJ 21941-972, Brazil}

\date{\today}

\begin{abstract}
We demonstrate the difference between local, single-particle dynamics and global dynamics of entangled quantum systems coupled to independent environments.  Using an all-optical
experimental setup, we show that,
while the environment-induced decay of each system is asymptotic, quantum 
entanglement may suddenly disappear. This ``sudden death"  constitutes yet another distinct and counter-intuitive trait of entanglement.     
\end{abstract}
\maketitle

 The real-world success of quantum
computation~\cite{chuang,preskill,bennett1} and
communication~\cite{bennett2,ekert,gisin,bennett3,bouwmeester,boschi,duan}
relies on the longevity of entanglement in multi-particle quantum
states. In this respect, the presence of decoherence~\cite{zurek} in communication
channels and computing devices, which stems from the unavoidable
interaction between these systems and the environment, presents a
considerable obstacle, as it degrades the entanglement when the
particles propagate or the computation evolves. Decoherence leads to
both local dynamics, associated with single-particle dissipation, diffusion, and
decay, and to global dynamics, which may provoke
the eventual disappearance of entanglement~\cite{diosi,dodd,yu1,franca,yu2}.   This phenomenon, known as ``entanglement sudden death" \cite{yu2}, is strikingly different from the single-particle dynamics, which occur asymptotically, and as a result has been the focus of much recent theoretical work \cite{diosi,dodd,yu1,franca,yu2}.    We have experimentally demonstrated the sudden death of entanglement of a two-qubit system under the influence of independent environments.  Our all-optical setup allows for the controlled investigation of a variety of dynamical maps that describe fundamental processes
in quantum mechanics and quantum information.
\par
Consider a two-level quantum system $S$ (upper and lower states $|e\rangle$
and $|g\rangle$, respectively) under the action of a zero-temperature reservoir $R$.
At zero temperature, the reservoir $R$ is in the $\ket{0}_R$
(vacuum) state, and the $S-R$ interaction can be represented by a
quantum map, known as the amplitude decay
channel~\cite{preskill,chuang}:
 \begin{equation}\label{qm}
\begin{array}{l}
 \left| g \right\rangle _S  \otimes \left| 0 \right\rangle _R
 \to \left| g \right\rangle _S  \otimes \left| 0 \right\rangle _R  \\
 \left| e \right\rangle _S  \otimes \left| 0 \right\rangle _R  \to
 \sqrt {1 - p} \left| e \right\rangle _S  \otimes \left| 0
 \right\rangle _R  + \sqrt p \left| g \right\rangle _S
 \otimes \left| 1 \right\rangle _R . \\
 \end{array}
\end{equation}
Under this map, the lower state $\ket{g}$ is not affected, while the
upper state $\ket{e}$ either decays to $\ket{g}$ with probability
$p$, creating one excitation in the environment (state
$|1\rangle_R$), or remains in $\ket{e}$, with probability $1-p$.
This would be the situation, for instance, in the spontaneous
emission of a two-level atom. In this case, the state $\left| 1
\right\rangle _R$ would correspond to one photon in the reservoir.
Under the Markovian approximation, $p = 1 - \exp \left( { - \Gamma
t} \right)$, that is, the decay probability approaches unity
exponentially in time. As an initial pure state
$a|e\rangle+b|g\rangle$ decays, it gets entangled with the
environment, gradually losing its coherence and its purity over
time. Complete decay only occurs asymptotically in time
($p\rightarrow1$ when $t\rightarrow \infty$), when the two-level
system is again described by the pure state $|g\rangle$.

Now consider two entangled qubits that decay according to the map
\eqref{qm}.  How does the entanglement of the two-qubit system
evolve?  Does it mimic the asymptotic decay of each qubit,
disappearing at $t\rightarrow\infty$, or does it disappear at some
finite time? This question has been explored
theoretically~\cite{diosi,dodd,yu1,franca,yu2}, but up to now there
has been no experimental investigation of the relation between the
global entanglement dynamics and the local decay of the constituent
subsystems.
\par
In order to adequately answer these questions, one needs a formal
definition of entanglement.  A convenient measure of entanglement
for a two-qubit system  is the {\it concurrence} $C$, introduced by
Wootters~\cite{wootters}, and given by
\begin{equation}\label{concurrence}
C = {\rm{max}}\left\{ {0,\Lambda} \right\}\,,
\end{equation}
where
\begin{equation}\label{lambda}
\Lambda=\sqrt {\lambda _1 }  - \sqrt {\lambda _2 } - \sqrt {\lambda
_3 }  - \sqrt {\lambda _4 }\,,
\end{equation}
and the quantities $\lambda _i$ are the positive eigenvalues, in
decreasing order, of the matrix
\begin{equation}
\rho \left( {\sigma _y  \otimes \sigma _y } \right)\rho ^* \left(
{\sigma _y  \otimes \sigma _y } \right)\,,
\end{equation}
where $\rho$ is the density matrix of the bipartite system. Here
$\sigma_y$  is the second Pauli matrix and the conjugation occurs in
the computational basis $\left\{ {\left| {00} \right\rangle ,\left|
{01} \right\rangle ,\left| {10} \right\rangle ,\left| {11}
\right\rangle } \right\}$. $C$ quantifies the amount of quantum
correlation that is present in the system, and can assume values
between 0 (only classical correlations) and 1 (maximal
entanglement).
\par
For the dynamics given by Eq.~(\ref{qm}), and an initial state of
the form $\ket{\Phi}=|\alpha| \left| {gg} \right\rangle +
|\beta|\exp(i\delta) \left| {ee} \right\rangle$, the entanglement
decay dynamics depends on the relation between $|\alpha|$ and
$|\beta|$~\cite{franca}. Concurrence in this case is given by
\begin{equation}\label{cinitial}
C={\rm max}\left\{0,2 \left( {1 - p} \right)\left| \beta
\right|\left( {\left| \alpha  \right| - p\left| \beta  \right|}
\right)\right\}\,.
\end{equation}
From this expression, one can see that for $\left| \beta \right| <
\left| \alpha \right|$, entanglement disappears only when the
individual qubits have completely decayed ($p = 1$), while for
$\left| \beta \right| > \left| \alpha  \right|$, entanglement
disappears for $p=|\alpha/\beta|<1$, which corresponds to a finite
time. This phenomenon has been called
``entanglement sudden death" \cite{yu2}.  Since the concurrence of the initial state $(p=0)$ is
$C=2|\alpha\beta|$, the entanglement dynamics of
two states with the same initial concurrence can be quite
different.
\par
Photons are a very useful experimental tool for demonstrating these
properties and, more generally, for studying quantum channels like
the one given in Eq.~(\ref{qm}), since the decoherence mechanisms
can be implemented in a controlled manner.  Let us associate the $H$
and $V$ polarizations of a photon respectively to the ground and
excited states of the two-level system $S$. The reservoir $R$ in
turn is represented by two different momentum modes of the photon.
\begin{figure}[b]
\centering
\includegraphics*[width=70mm]{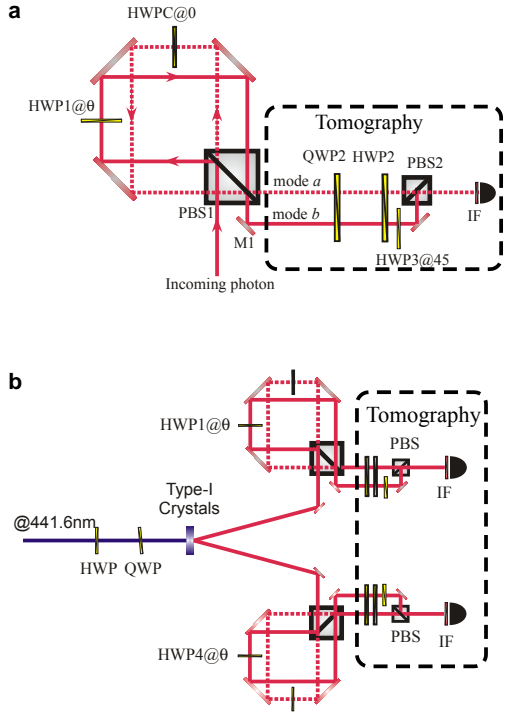}
\caption{Experimental setup. {\bf a} Amplitude decay channel for a single photonic qubit.
Due to the polarized beam splitter PBS1, the $H$ and $V$ components
of polarized photons propagate respectively along counter-clockwise
and clockwise paths within the interferometer.  With half-wave
plate HWP1 set at $0^\circ$, they are coherently recombined into the
outgoing spatial mode $a$, which represents the ``vacuum state'' of
the reservoir. For other angles of HWP1, the $V$-component undergoes
a rotation, corresponding to its probabilistic ``decay'' into the
$H$-component, which PBS1 sends to outgoing mode $b$, representing
the ``one-excitation state'' of the reservoir. Wave plates HWP2 and
QWP2, together with PBS2, are used for tomography of the
polarization state of outgoing modes $a$ and $b$, which are
recombined incoherently on PBS2 using HWP3, and sent to the same
detector.  IF is an interference filter.   {\bf b}   Amplitude decay channel applied to entangled qubits.  The entangled state is generated by parametric
down-conversion in type-I non-linear crystals.}\label{setupone}
\end{figure}
Fig. \ref{setupone}a shows a Sagnac-like interferometer which implements the
amplitude-decay channel \eqref{qm} for a single qubit.   A photon,
initially in the incoming part of mode $a$, is split into its
horizontal ($H$) and vertical ($V$) polarization components by a
polarizing beam splitter (PBS1). Let us ignore the half-wave plates
HWP1 and HWPC momentarily. The $V$-polarization component is
reflected and propagates through the interferometer in the clockwise
direction, and, if unaltered, reflects through PBS1 into the
outgoing part of mode $a$. The $H$-polarization component is
transmitted and propagates through the interferometer in the
counter-clockwise direction and transmits through PBS1, also into
the outgoing part of mode $a$.  However, the interferometer is aligned so that the two
paths are spatially separated, making it possible to manipulate the
$H$ and $V$ polarization components independently.

To realize the amplitude decay given in map \eqref{qm}, we use HWP1
to rotate the polarization of the $V$ component to
$\cos(2\theta)\ket{V}+\sin(2\theta)\ket{H}$, where $\theta$ is the
angle of HWP1. Suppose that an incoming photon is $V$-polarized.
When this photon exits the interferometer through PBS1, it is
transmitted into mode $b$ with probability $p=\sin^2(2\theta)$ and
reflected into mode $a$ with probability $\cos^2(2\theta)$.  This
evolution can thus be described by $\ket{V}\ket{a}\longrightarrow
\sqrt{1-p}\ket{V}\ket{a} + \sqrt{p} \ket{H}\ket{b}.$ Identifying the
outgoing modes $a$ and $b$ (which correspond to orthogonal spatial
modes) as the states of the reservoir with zero and one excitation,
respectively, this operation is equivalent to that on the
$\ket{e}\ket{0}_R$ state in map \eqref{qm}.  An incoming
$H$-polarized photon is left untouched, corresponding to the first
line in Eq.~\eqref{qm}. This process is therefore identical to the
decay of a two-level system. Half-wave plate HWPC, oriented at
$0^\circ$, is used solely to match the lengths of the two optical
paths.    The path lengths are adjusted so that if HWP1 is oriented
at $0^\circ$, the polarization state in mode $a$ after the
interferometer is exactly the same as the input state. Photons in
modes $a$ and $b$ are then directed to the same quantum state
tomography (QST) system, composed of a quarter-wave plate QWP2,
half-wave plate HWP2, and the polarizing beam splitter PBS2, and
then registered using a single-photon detector, equipped with a 10
nm FWHM interference filter, and a $1.5$ mm diameter aperture. Mode
$b$ is recombined incoherently with mode $a$ on PBS2, so that both
modes can be detected with a single detector. This is achieved by
assuring that the path length difference between modes $a$ and $b$
is greater than the coherence length of the photons, which is
determined by the width of the interference filters ($\sim
0.1$\,mm). The half-wave plate HWP3, aligned at $45^\circ$,
transforms $H$-polarized photons into $V$-polarized ones.  Since
PBS2 transmits $H$-polarization and reflects $V$-polarization, the
combination of HWP3 and PBS2 reflects photons that were originally
$H$-polarized. This assures that the QST performed is identical for
both modes $a$ and $b$.
\par
Using the interferometer described above, we studied both the decay
of a single-qubit and the dynamics of two entangled two-level
systems interacting with independent amplitude-decay reservoirs. The
experimental setup is shown in Fig.~\ref{setupone}b.
Polarization-entangled photon pairs with wavelength centered around
884 nm were produced using a standard source \cite{kwiat99} composed
of two adjacent type-I LiIO$_3$ nonlinear crystals pumped by a 441.6
nm c.w. He-Cd laser. One crystal produces photon pairs with
$V$-polarization and the other produces pairs with $H$-polarization.
After propagation and spatial mode filtering, the $H$ and $V$ modes
are spatially indistinguishable, and a photon pair is described by
the pure state $\ket{\Phi}=|\alpha|\ket{HH}+|\beta|
e^{\textrm{i}\delta}\ket{VV}$ with high fidelity. A half-wave plate
(HWP) and a quarter-wave plate (QWP), placed in the pump beam, allow
the control of the coefficients $|\alpha|$ and $|\beta|$, and the
relative phase $\delta$ of the state \cite{kwiat99}.
\par
\begin{figure}[t]
\centering
\includegraphics*[width=70mm]{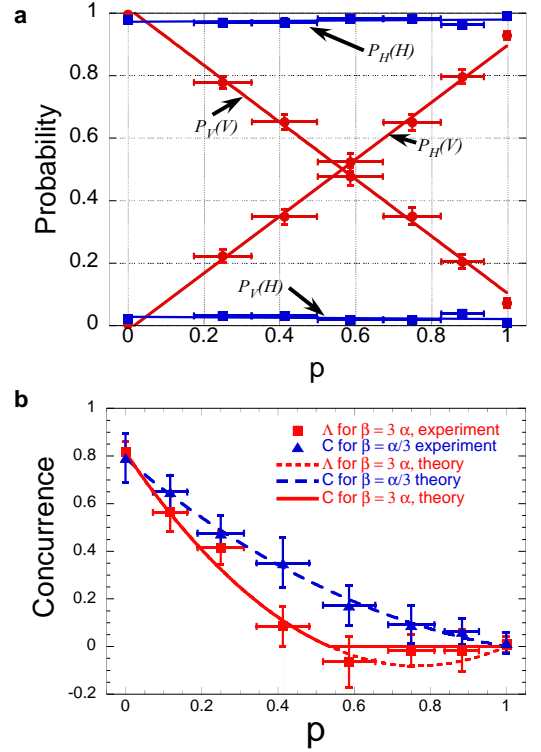}
\caption{Results for amplitude decay channel.  {\bf a} Experimental amplitude decay for a  single qubit.
$P_{V}(V)$ and $P_{H}(V)$ are the probabilities of detecting an
input $V$-polarized photon in the $V$ and $H$ states, respectively.
$P_{V}(H)$ $P_{H}(H)$ are the the probabilities for an input $H$
photon.  The points correspond to experimental data, and the lines
are linear fits. \label{result2}
{\bf b} Entanglement decay as a function of the probability $p$.
The squares correspond to experimentally obtained values of
$\Lambda$ for the case $|\beta|^2 = 3|\alpha|^2$.  The solid line is
the theoretical prediction of the concurrence for this state, given
by Eq. \ref{concurrence}, while the dotted line shows the value of
$\Lambda$, given by Eq. \eqref{lambda}.  The triangles are
experimental values of $\Lambda$ for the case $|\beta|^2 =
|\alpha|^2/3$, and the dashed line is the theoretical prediction for
$\Lambda$ and $C$, which are equivalent for this state.
\label{result1}}
\end{figure}

The decay of a single qubit was investigated experimentally for both
$H$ and $V$-polarized photons, by generating states $|VV\rangle$ and
$|HH\rangle$, and registering coincidence counts, with one photon
propagating through the interferometer and the other serving as a
trigger. The coincidence detection window
($c\times5\,\mathrm{ns}\sim1.5$\, m) was larger than the path
difference between outgoing modes $a$ and $b$ ($\sim5$ cm).
Fig.~\ref{result1}a shows $P_{V}(V)$, $P_{H}(V)$, $P_{V}(H)$
$P_{H}(H)$ as a function of $p$, where $P_{J}(K)$ is probability of
finding an input $K$-polarized photon in the $J$ state after the
interferometer.  The linear behavior in $p$ is characteristic of
exponential decay in $t$, given that $p = 1 - \exp \left( { - \Gamma
t} \right)$. Vertical error bars were obtained by standard Monte
Carlo simulations \cite{altepeter}. The horizontal error bars
represent uncertainty in aligning the waveplates.

For the investigation of the entanglement dynamics, non-maximally
entangled states were produced, and each photon sent to a separate
interferometer, which implemented an amplitude-damping reservoir,
and then to a QST system.   The half-wave plates HWP1 and HWP4 were
set to the same angle $\theta$, so that the reservoirs, though
independent, acted with the same probability $p$.  QST of the output
two-photon state followed the usual recipe consisting of a set of
sixteen coincidence measurements \cite{kwiat-tomo}.  We repeated the
same procedure for different values of $p$, obtaining the
tomographic reconstruction of the output two-photon polarization
state in all cases.  The concurrence was calculated using Eqs.
(\ref{concurrence}) and (\ref{lambda}).

\par
Figure~\ref{result1}b displays the concurrence, and the quantity
$\Lambda$, given by Eq.~(\ref{lambda}), as a function of the decay
probability $p$, for two initial states that, although not pure, are
very close to $\ket{\Phi}=|\alpha|\ket{HH}+|\beta|
e^{\textrm{i}\delta}\ket{VV}$: state I, defined by $|\beta|^2 =
|\alpha|^2/3$ (triangles), and state II, defined $|\beta|^2 =
3|\alpha|^2$ (squares).   Tomography of the initial states I and II
showed them to have the same concurrence ($\sim0.8$), and similar
purity ($\sim0.91-0.97$).  The theoretical curves were obtained by
applying map (\ref{qm}) to the experimentally determined initial
states, corresponding to $p=0$.  As before, the vertical error bars
were obtained by Monte Carlo simulation \cite{altepeter}. For
initial state I, entanglement disappears asymptotically, and the
concurrence goes to zero only when both individual systems have
decayed completely ($p=1$). For initial state II, however, the
entanglement behaves very differently: the concurrence goes to zero
at a finite time ($p < 1$).  This demonstrates ``entanglement sudden
death.''

\begin{figure}[t]
\begin{center}
\includegraphics*[width=70mm]{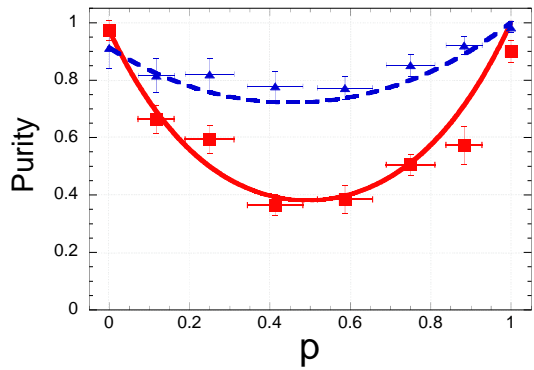}
\caption{Purity as a function of $p$ for the amplitude damping
channel.       The squares correspond to experimentally obtained
values of the purity for the case $|\beta|^2 = 3|\alpha|^2$, while
the solid line is the theoretical prediction.  The triangles are
experimental values of the purity for the case $|\beta|^2 =
|\alpha|^2/3$, and the dashed line is the corresponding theoretical prediction.
\label{result1b}}
\end{center}
\end{figure}

It is also illustrative to study the purity, defined as
$\mathrm{tr}\rho^2$, in function of the decay probability, as shown
in Fig. \ref{result1b} for states I and II. We see that in both
cases the purity reaches a minimum but is restored when $p = 1$,
when all photons have ``decayed" to the $H$-polarization state.

\begin{figure}[b]
\centering
\includegraphics*[width=70mm]{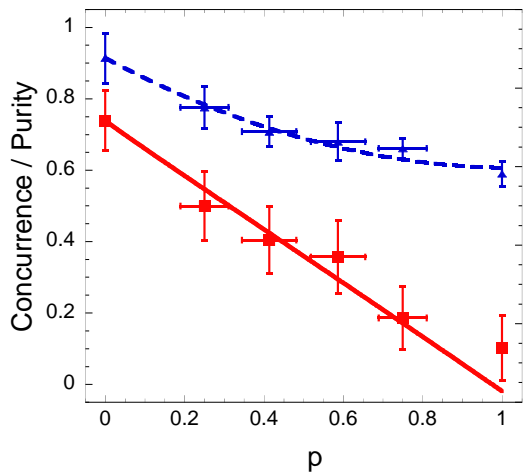}
\caption{Experimental results for the dephasing reservoir.   Concurrence (squares) and purity (triangles) are shown for the case
$|\beta|^2=3|\alpha|^2$.  The solid line is the corresponding theoretical prediction for concurrence, given by Eq. \eqref{concurrence}.  The dashed line is the theoretical prediction for purity, given by $\mathrm{tr}\rho^2$.  The concurrence goes to zero
asymptotically. \label{result3}}
\end{figure}
\par
In order to further illustrate the usefulness of the present scheme
for studying decoherence of entangled systems, we have performed a
second experiment studying the action of a pair of dephasing
reservoirs, described by the map \cite{preskill,chuang}
 \begin{equation}\label{qm2}
\begin{array}{l}
 \left| g \right\rangle _S  \otimes \left| 0 \right\rangle _R
 \to \left| g \right\rangle _S  \otimes \left| 0 \right\rangle _R  \\
 \left| e \right\rangle _S  \otimes \left| 0 \right\rangle _R  \to
 \sqrt {1 - p} \left| e \right\rangle _S  \otimes \left| 0
 \right\rangle _R  + \sqrt p \left| e \right\rangle _S
 \otimes \left| 1 \right\rangle _R . \\
 \end{array}
\end{equation}
This map represents elastic scattering between atom and reservoir.
States $|e\rangle$ and $|g\rangle$ are not changed by the
interaction, but any coherent superposition of them gets entangled
with the reservoir. There is no longer decay, but only loss of
coherence between ground and excited states. The dephasing map can
be implemented with the same interferometer through the addition of
an extra HWP at $45^o$ in mode $b$ before the QST system (or,
equivalently, through the removal of HWP3 and redefinition of the
QST  measurements).  For the dephasing channel, pure
states I and II present identical behavior, becoming completely
disentangled only when $p=1$. The concurrence (squares) and bipartite purity (triangle)
as a function of $p$ for the entangled state II is shown in Fig.
\ref{result3}. The theoretical prediction for the concurrence
disappears slightly before $p=1$, due to the fact that the initial
state is not 100\% pure.
\par
In conclusion, we have reported an experimental demonstration of the
sudden disappearance of the entanglement of a bipartite system,
induced by the interaction with an environment.   We have shown that
entangled states with the same initial concurrence may exhibit, for
the same reservoir, either an abrupt or an asymptotic disappearance
of entanglement, in spite of the fact that the constituents of the
system always exhibit an asymptotic decay. We have explicitly
demonstrated that this behavior also depends on the characteristics
of the reservoir, through two examples, corresponding to amplitude
decay and dephasing.  The experimental setup  represents a reliable
and simple method for studying decoherence of entangled systems
interacting with controlled reservoirs.
\begin{acknowledgments}
The authors acknowledge financial support from the Brazilian funding
agencies CNPq, CAPES, PRONEX, FUJB and FAPERJ. This work was
performed as part of the Brazilian Millennium Institute for Quantum
Information.
\end{acknowledgments}



\begin{thebibliography}{99}



\bibitem{chuang} Nielsen, M. A. \& Chuang I, {\it Quantum Computation and
Quantum Information} (Cambridge University Press, Cambridge, U.K.,
2000).

\bibitem{preskill} Preskill, J. {\it Lecture notes on quantum
information and computation}. Available at
http://www.theory.caltech.edu/people/preskill/ph219/.

\bibitem{bennett1} Bennett, C. H. \& DiVincenzo, D. P. Quantum
information and computation. {\it Nature} {\bf 404}, 247-255 (2000).

\bibitem{bennett2} Bennett, C. H. \& Brassard, G. in {\it Proc. IEEE Int.
Conf. on Computers, Systems and Signal Processing} 175-179 (IEEE,
New York, 1984).

\bibitem{ekert} Ekert, A. Quantum cryptography based on Bell's theorem.
{\it Phys. Rev. Lett.} {\bf 67}, 661-663 (1991).

\bibitem{gisin} Gisin, N. {\it et al.}, Quantum Cryptography,
{\it Rev. Mod. Phys.} {\bf 74} 145 (2002).

\bibitem{bennett3} Bennett, C. H. {\it et al}. Teleporting an unknown
quantum state via dual classical and Einstein-Podolsky- Rosen
channels. {\it Phys. Rev. Lett.} {\bf 70}, 1895-1898 (1993).

\bibitem{bouwmeester} Bouwmeester, D. et al. Experimental quantum
teleportation. {\it Nature} {\bf 390}, 575-579 (1997).

\bibitem{boschi} Boschi D. {\it et al}. Experimental Realization of
Teleportating an Unknown Pure Quantum State via Dual Classical and
Einstein-Podolsky-Rosen Channels. {\it Phys. Rev. Lett} {\bf 80},
1121 (1998).

\bibitem{duan} Duan, L.-M., Lukin, M. D., Cirac, J. I., \& Zoller,
P.  Long-distance quantum communication with atomic ensembles and
linear optics. {\it Nature} {\bf 414}, 413-418 (2001).

\bibitem{zurek} Zurek, W. H. Decoherence, einselection, and
the quantum origins of the classical. {\it Rev. Mod. Phys.} {\bf
75}, 715-775 (2003).

\bibitem{diosi} Di\'osi, L.  in {\it Irreversible Quantum Dynamics},
edited by F. Benatti
and R. Floreanini (Springer, Berlin, 2003).

\bibitem{dodd} P. J. Dodd, P. J. \& Halliwell, J. J. Disentanglement
and decoherence by open system dynamics. {\it Phys. Rev. A} {\bf 69}, 052105
(2004).

\bibitem{yu1} Yu, T. \& Eberly, J. H. Finite-time disentanglement via
spontaneous emission. {\it Phys. Rev. Lett.} {\bf 93}, 140404
(2004).

\bibitem{franca} Santos, M. F., Milman, P., Davidovich, L. \& Zagury, N.,
Direct measurement of finite-time disentanglement induced by a reservoir,
{\it Phys. Rev. A} {\bf 73}, 040305(R) (2006).

\bibitem{yu2} Yu, T \& Eberly, J. H. Quantum open system theory: Bipartite aspects.
{\it Phys. Rev. Lett.} {\bf 97}, 140403 (2006).


\bibitem{wootters} Wootters, W. K. Entanglement of formation of an
arbitrary state of two qubits. {\it Phys. Rev. Lett.} {\bf 80}, 2245-2248 (1998).

\bibitem{kwiat99}  Kwiat, P. G.  {\it et al.}, Ultrabright source of
polarization-entangled photons, {\it Phys. Rev. A.} {\bf 60} R773 (1999).

\bibitem{kwiat-tomo} James D. V. {\it et al.}, Measurement of qubits,
{\it Phys. Rev. A.} {\bf 64} 052312 (2001).

\bibitem{altepeter} Altepeter, J. B. {\it et al.}, Photonic State Tomography,
Advances in Atomic, Molecular and Optical Physics, Elsevier (2005).


\end{thebibliography}
\end{document}